\documentstyle[twoside]{article}
\newcount\Mac  \Mac=0  
\newcommand{\ifMac}[2]{\ifnum\Mac=1 #1 \else #2 \fi}

\newcommand{\Tr}{\mathop{\rm Tr}}

\newcommand{\GeV}{\,{\rm GeV}}

\newcommand{\NP}{Nucl. Phys.}
\newcommand{\PRL}{Phys. Rev. Lett.}
\newcommand{\PL}{Phys. Lett.}
\newcommand{\PR}{Phys. Rev.}
\newcommand{\R}{{\cal R}}
\newcommand{\mhu}{m_{h^{\rm u}}}
\newcommand{\mhd}{m_{h^{\rm d}}}
 
\newcommand{\muR}{m_{\tilde{u}_R}} \newcommand{\muL}{m_{\tilde{u}_L}}
 
\newcommand{\mdR}{m_{\tilde{d}_R}} 

\newcommand{\meR}{m_{\tilde{e}_R}} 
\newcommand{\mL}{m_{\tilde{L}}}  \newcommand{\mQ}{m_{\tilde{Q}}}

\newcommand{\eq}[1]{~{\rm (\ref{eq:#1})}}
\newcommand{\sys}[1]{~{\rm (\ref{sys:#1})}}

\newcommand{\MGUT}{M_{\rm U}}
\newcommand{\MM}{M_{\rm M}}
\newcommand{\MF}{M_{\rm F}}
\def\Red{}
\def\Black{}
\def\Blue{}

\newcommand{\lascia}[1]{}
\makeatletter
\def\art{\@ifnextchar[{\eart}{\oart}}
\def\eart[#1]#2#3#4#5#6{{\rm #2}, {\em #3 \bf #4} {\rm (#6) #5}}
\def\hepart[#1]#2{{\rm #2, \em#1}}
\newcommand{\oart}[5]{{\rm #1}, {\em #2 \bf #3} {\rm (#5) #4}}
\newcommand{\y}{{\rm and} }

\newcounter{alphaequation}[equation]
\def\thealphaequation{\theequation\hbox to
0.6em{\hfil\alph{alphaequation}\hfil}}
\def\eqnsystem#1{
\def\@eqnnum{{\rm (\thealphaequation)}}
\def\@@eqncr{\let\@tempa\relax \ifcase\@eqcnt \def\@tempa{& & &} \or
  \def\@tempa{& &}\or \def\@tempa{&}\fi\@tempa
  \if@eqnsw\@eqnnum\refstepcounter{alphaequation}\fi
\global\@eqnswtrue\global\@eqcnt=0\cr}
\refstepcounter{equation} \let\@currentlabel\theequation \def\@tempb{#1}
\ifx\@tempb\empty\else\label{#1}\fi
\refstepcounter{alphaequation}
\let\@currentlabel\thealphaequation
\global\@eqnswtrue\global\@eqcnt=0 \tabskip\@centering\let\\=\@eqncr
$$\halign to \displaywidth\bgroup \@eqnsel\hskip\@centering
$\displaystyle\tabskip\z@{##}$&\global\@eqcnt\@ne
\hskip2\arraycolsep\hfil${##}$\hfil& \global\@eqcnt\tw@\hskip2\arraycolsep
$\displaystyle\tabskip\z@{##}$\hfil
\tabskip\@centering&\llap{##}\tabskip\z@\cr}

\def\endeqnsystem{\@@eqncr\egroup$$\global\@ignoretrue} \makeatother

\def\SU{{\rm SU}}
\def\circa#1{\,\raise.3ex\hbox{$#1$\kern-.75em\lower1ex\hbox{$\sim$}}\,}

\begin{document}
\twocolumn[
\begin{quote}{\em 18 May 1997\hfill \bf hep-ph/9705306}\end{quote}
 \vspace{1cm}
\centerline{\huge\bf\Red Distinguishing gauge-mediated}
\centerline{\huge\bf from unified-supergravity spectra}

\bigskip\bigskip\Black
\centerline{\large\bf Alessandro Strumia} \vspace{0.3cm}
\centerline{\em Dipartimento di Fisica, Universit\`a di Pisa {\rm and}}
\centerline{\em INFN, sezione di Pisa,  I-56126 Pisa, Italia}\vspace{0.8cm}
\Blue

\centerline{\large\bf Abstract}
\begin{quote}\large\indent
We show that gauge-mediation and unified-supergravity
give sufficiently firm and different predictions for the spectrum
of supersymmetric particles
to make it possible to discriminate the two scenarios
even if the messenger mass is close to the unification scale.
\end{quote}\Black
\vspace{1cm}]

\paragraph{1}
According to our present theoretical understanding,
the presence of supersymmetric partners of the
three generation of fermions 
has a very different impact on flavour physics,
depending on the relative order
of few fundamental high-energy scales.
If the hardness scale of the supersymmetry-breaking
soft terms (`{\em mediation scale}', $\MM$)
is higher than the hardness scale of
the standard Yukawa couplings (`{\em flavour scale}', $\MF$)
or of the {\em unification scale}
$\MGUT\approx 2\cdot 10^{16} \GeV$,
we expect that the sfermion mass matrices contain new
sources of flavour and CP violation, most
likely detectable due to the heavyness of the top quark~\cite{FVGUT}.
In this case it is quite possible that the consequent effects,
due to virtual sparticles exchanges,
will be discovered even before than the sparticles themselves.
If instead the supersymmetry-breaking
soft terms are mediated at a lower scale where
the flavour and unification physics have decoupled, we
expect that the only flavour violation present at low energies
is described by the supersymmetrized extension of the standard CKM matrix.
In this case supersymmetric loops could give non negligible
contributions only to `standard' flavour and/or CP violating effects,
mainly to the $b\to s\gamma$ and $b\to s\ell^+\ell^-$ decays~\cite{FVMSSM}  .

If this view is correct, the forthcoming experiments about flavour physics
should either discover some signal (or combination of signals), thus
giving a strong hint in favour of the first scenario, or
exclude new flavour and CP violations up to a certain level,
making the first scenario less interesting.
In any event, it is clearly useful to have an alternative way of discriminating
between the two scenarios.
If the mediation scale is sufficiently low ($\MM\circa{<}10^8\GeV$)
the decay  within the detector
of the lightest supersymmetric particle (LSP) into a gravitino
would give such an incontrovertible signal.
No such a clean signal is present in the remaining range of $\MM$.

\smallskip

In this paper we want to show that, within reasonably minimal models, the spectrum
of the supersymmetric particles typical of the
two scenarios is sufficiently different that is possible to recognize which
of the two scenarios is actually realized,
even if the mediation scale gets close to the unification scale.

\begin{figure*}[t]\setlength{\unitlength}{1cm}
\begin{center}\begin{picture}(15,9)
\ifMac{\put(-0.5,0){\special{picture 5}}}
{\put(-0.5,0){\includegraphics{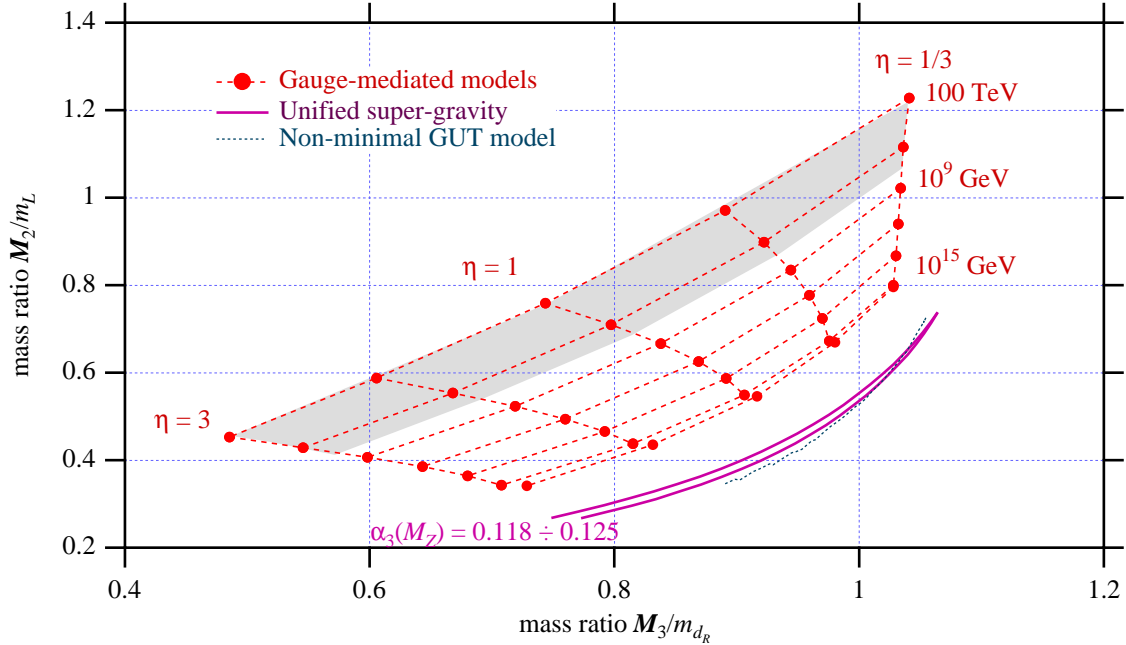}}}
\end{picture}
\caption[1]{\em Correlations $(M_3/\mdR,M_2/\mL)$
as predicted by unified supergravity (continuous line)
and gauge-mediation (dashed lines) for different values
of the messenger scale and of the parameter $\eta$, defined in\sys{GM}.
We have employed $\alpha_3(M_Z)=0.118$.
The supergravity
and gauge-mediation predictions with $\MM=10^{15}\GeV$ are also
shown for $\alpha_3(M_Z)=0.125$ (lower lines).
In the shaded area it is possible to observe the clean signature of LSP decay.
The dotted line refers to a non minimal unification model,
as discussed in the text.}
\end{center}\end{figure*}

\begin{figure*}[t]\setlength{\unitlength}{1cm}
\begin{center}\begin{picture}(15,9)
\ifMac{\put(-0.5,0){\special{picture 10}}}
{\put(-0.5,0){\includegraphics{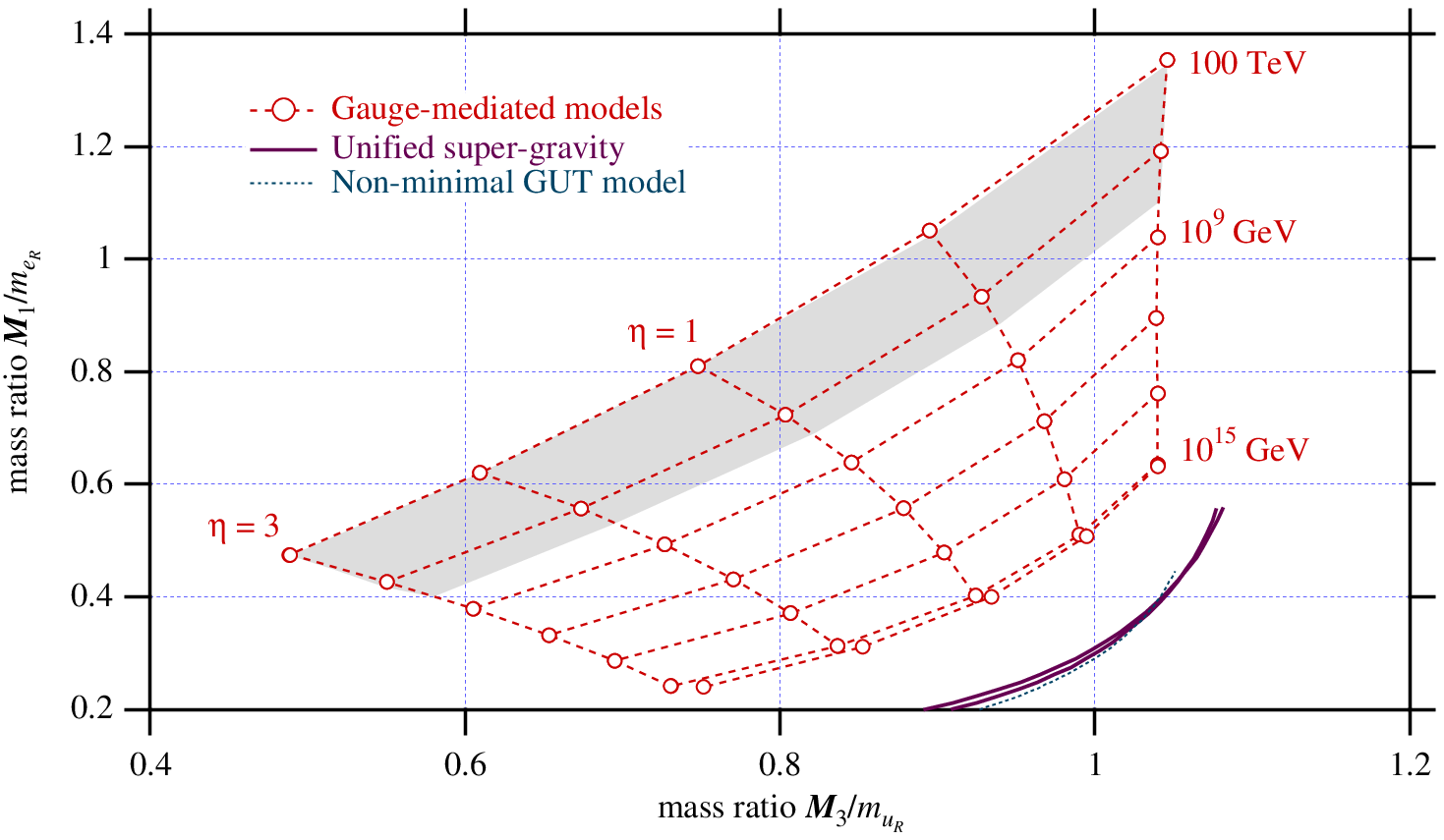}}}
\end{picture}
\caption{\em As in fig.~1, except that we plot the
correlations $(M_3/\muR,M_2/\meR)$ involving masses
of sparticles unified in a $10$ of $\SU(5)$.}
\end{center}
\end{figure*}

\paragraph{2}
Both the two scenarios outlined above can be realized in a
clean and predictive way.
The first case, $\MM\circa{>}\min(\MF,\MGUT)$,
arises naturally if supergravity interactions
mediate the soft terms~\cite{SuGraSoft}.
In this case the hardness scale of the soft terms, $\MM$,
is the reduced Planck mass.
Low energy physics suggests that the field theory at this high scale
has a unified gauge group,
$G\supseteq\SU(5)\supset\SU(3)_{\rm c}\otimes\SU(2)_L\otimes{\rm U}(1)_Y$,
so that the sparticle masses are subject to unification relations.
Neglecting, for the moment, SU(5)-breaking effects,
unified supergravity predicts
\begin{eqnsystem}{sys:SuGra}\label{eq:MiGUT}
M_5 &=& M_i(\MGUT)\\ 
\label{eq:5}  m_{\bar{5}} &=& \mL(\MGUT)=\mdR(\MGUT)\\
\label{eq:10}      m_{10} &=& \meR(\MGUT)=\mQ(\MGUT)=\muL(\MGUT)
\end{eqnsystem}
where $M_i(\MGUT)$ are the three gaugino mass parameters
($i=1,2,3$ runs over the three factors of the SM gauge group)
and $m_{\tilde{f}}(\MGUT)$ are the soft mass terms for the sfermions $\tilde{f}$,
all renormalized at the unification scale $\MGUT$.
Unacceptably large flavour violations are avoided if
the sparticles of the first generation
are degenerate with the corresponding ones of the second generation
at a level that, for our present purposes,
can be considered perfect.
For the purposes of this paper,
the third generation sparticles are less interesting because
their unification relations
are easily broken~\cite{SoglieGUT} and corrected by not sufficiently
controlled RGE effects~\cite{RGE}.

\bigskip

Also the second scenario,
$\MM<\min(\MGUT,\MF)$, arises naturally if the soft terms
are communicated by the standard gauge interactions
at some `mediation scale' $\MM$, lower than the unification scale\footnote{
If the gauge group at this scale is unified, or partially unified,
new sources of flavour and CP violations are again
present in the supersymmetry-breaking soft terms.},
where some charged `messenger' superfields
feel directly the breaking of supersymmetry~\cite{GaugeSoft}.
In this case the spectrum of the supersymmetric particles
is mainly determined by their gauge charges.
More precisely, in a large class of minimal models the prediction
for the soft terms, at the messenger mass $\MM$, can be
conveniently parametrized as
\begin{eqnsystem}{sys:GM}
M_i(\MM)&=& \frac{\alpha_i(\MM)}{4\pi} M_0,\\
m_R^2(\MM) &=&  \eta\cdot c^i_R M_i^2(\MM),
\end{eqnsystem}
where 
$m_R$ are the soft mass terms for the fields
$$R=\tilde{Q},\tilde{u}_R,\tilde{d}_R,\tilde{e}_R,
\tilde{L},h^{\rm u},h^{\rm d},$$
and the various quadratic Casimir coefficients $c^i_R$ are
listed in table~\ref{Tab:bcude}.
Here $M_0$ is an overall mass scale and $\eta$
parametrizes the different minimal models.
For example $\eta=(n_5+3n_{10})^{-1/2}$ in models
where a gauge singlet couples supersymmetry breaking
to $n_5$ copies of messenger fields in the $5\oplus\bar5$
representation of SU(5)
and to $n_{10}$ copies in the $10\oplus\overline{10}$
representation~\cite{GaugeSoft,DTW,BMPZ}.
Values of $\eta$ bigger than one are possible
if more than one supersymmetry-breaking singlet is present,
since an $R$-symmetry can suppress the gaugino masses
with respect to the scalar masses~\cite{GaugeSoft,DTW,BMPZ}.
We will not employ the additional prediction
that the $A$-terms vanish at the messenger scale.

Gauge-mediation models have the generic problem that
gauge interactions alone cannot mediate
the `$\mu$-term',
as well as the corresponding `$B\cdot \mu$-term',
since these terms break a Peccei-Quinn symmetry.
Since the unknown physics
required to solve this problem
may easily give rise to unknown non minimal
contributions to the soft terms in the Higgs sector~\cite{GMmu},
we will concentrate on the safer
predictions for sfermion masses.

\paragraph{3}
The two scenarios do not differ in their predictions
for gaugino masses.
We now show that these scenarios
give instead sufficiently different predictions
for some scalar masses\footnote{Similar analyses
have been done in ref.~\cite{BMPZ,Peskin} under stronger
assumptions, like a universal supergravity spectrum at the unification scale
or a low messenger mass, and
neglecting non minimal contributions to the gauge-mediated spectrum.}.
The main feature at the basis of this fact is that,
even for a messenger scale close to the unification scale,
only the MSSM gauge bosons are relevant
for mediating supersymmetry breaking.
As a consequence, the Casimir coefficients
are different for particles coming from the same
unified representation, so that
the gauge-mediated spectrum is not unified
even when the messenger mass is close to the unification scale.

To illustrate this fact in approximate analytic form,
we recall the well known relations
between the soft terms at the Fermi scale
(as obtained by one loop RGE rescaling~\cite{RGE} down to $Q\approx 400\GeV$)
predicted by unified supergravity~\cite{MassRels}:
\begin{eqnsystem}{sys:SumSuGra}
\mdR^2-0.69 M_3^2&=&\mL^2,\\
\muR^2-0.75 M_3^2&=&\meR^2,
\end{eqnsystem}
with $M_2\approx 2 M_1\approx 0.3 M_3$.
A third less useful relation could be used
to substitute the $\tilde{u}_R$ squark
with a $\tilde{u}_L$ or $\tilde{d}_L$ one.

By similar rescaling, from the initial condition\sys{GM},
the corresponding prediction of gauge mediated models
are quite different,
even if the messenger scale is very high, $\MM=10^{15}\GeV$:
\begin{eqnsystem}{sys:SumGM}
\mdR^2-0.63 M_3^2&=&1.9\mL^2,\\
\muR^2-0.73 M_3^2&=&4.8\meR^2.
\end{eqnsystem}
In order to be more general,
to include subdominant corrections and
to give a more intuitive view of their difference,
we compare the predictions of the two scenarios in figures~1 and~2
in the full parameter space.
In figure~1 we show the different predictions
of the two scenarios for the combined mass ratios
$(M_3/\mdR,M_2/\mL)$ involving sfermions unified in a $\bar{5}$ of SU(5).
In figure~2 we show the mass ratios
$(M_3/\muR,M_1/\meR)$ involving sfermions unified in a 10 of SU(5).
The gaugino mass parameters $M_i$ are of course again related among themselves
and we have varied the ratio between gaugino and scalar masses
in a reasonable range\footnote{In the supergravity case the
sfermions of first and second generation could
be much heavier than the other sparticles~\cite{FTGUT}.
This feature alone would allow to recognize the scenario,
so that our plots do not cover this case.}, as indicated in the plots.
As anticipated, the difference between the predictions
is not unobservably small, even for $\MM=10^{15}\GeV$.

In our plots we have included 2-loop RGE effects~\cite{RGE2loop} and
threshold corrections at the electroweak scale, computed
in logarithmic approximation~\cite{SoglieMZ}.
These effects give a shift of the predictions that, in the plots,
is at most $20\%$ of the `distance' between the two more similar cases
of supergravity and gauge-mediation with $\MM=10^{15}\GeV$.
Furthermore, the shift is almost equal for the two analogous spectra.

We have preferred to plot the
gaugino and scalar mass parameters
commonly used to parametrize the sparticle spectrum
rather than the physical masses of the same particles.
With this choice, each plot covers
practically the fully general dependence on all
the supersymmetric parameters, \hbox{i.e.} the predictions are
practically independent on the other supersymmetric parameters
that do not appear in the plots
(the $\mu$ term, the overall scale of supersymmetric particles, $\tan\beta$, etc.).
More precisely, these variables affect our plots only through
two loop RGE effects and threshold corrections.
Since it is conceivable that also these parameters will be measured,
we have not included the relatively
small uncertainties associated
with their variations.
Rather we have assigned fixed reasonable values to these parameters,
as we now discuss in some detail:
\begin{itemize}
\item As said, we have not plotted sparticle pole masses.
For example $\muR$ is the soft term $\muR(\muR)$.
To compute the physical masses one has to
add the well known
$\SU(2)_L\otimes{\rm U}(1)_Y$-breaking effects,
that depend on $\tan\beta$,
and the known full one-loop effects,
that give up to 10\% corrections~\cite{1loopSoft}.
With this choice our plots are valid for
any moderate value of $\tan\beta$\
(we have used $\tan\beta=2$)\footnote{A small dependence of our plots
on $\tan\beta$ arises due to a two-loop RGE correction
$\delta m_R^2\propto Y_R g_1^2\lambda_t^2$
($\lambda_t=m_t/(v\sin\beta)$).}.

\item 
The overall scale of sparticle masses has been fixed
by choosing $M_3=500\GeV$.
Any other reasonable value of this measurable parameter gives
a shift comparable to the uncertainty
associated with the experimental error on $\alpha_3$,
as discussed below.

\item
In the supergravity case, the mass ratios in fig.s~1, 2
depend on the ratio
$m_{\bar{5}}/m_{10}$ only via numerically irrelevant
two loop and threshold effects at the electroweak scale.
In any case we have chosen $m_{\bar{5}}\approx 1.3 m_{10}$
in order to obtain a slepton mass ratio similar to the one
predicted by the high-$\MM$ gauge-mediated scenario.

Moreover, we have assumed that the
(again irrelevant) masses of third generation sfermions
are, at the unification scale, $50\%$ smaller than the
corresponding ones of the other generation,
as suggested by RGE effects above the unification scale.

We have set all
the $A$-terms at zero at the unification scale.
In principle, only the stop $A$ term, $A_t$, could be relevant for our purposes.
In practice, the induced effects (via two loop RGE and via threshold effects)
are again negligible\footnote{
Even a very large $A_t$ does not induce significant effects,
because $A_t$ is driven towards its infrared fixed point,
where $A_t\approx 2 M_2$.}.

\item In the gauge-mediation case,
we have neglected (computable) `threshold' corrections at
the messenger scale.
We expect that these corrections be important 
only in the case of light messengers,
i.e.\ when the difference with the unified supergravity scenario is large
anyhow.

\begin{table}[t]
$$\begin{array}{|c|ccccc|}\hline
i&c_i^Q&c_i^{u_R}&c_i^{d_R}&c_i^L&c_i^{e_R}\\[0.5mm] \hline
1 &{1\over30}&{8\over15}&{2\over15}&\vphantom{X^{X^X}}
{3\over10}&{6\over5}\\
2&{3\over2}&0&0&{3\over2}&0\\
3&{8\over3}&{8\over3}&{8\over3}&0&0\\ \hline
\end{array}$$
\caption{\em Values of the Casimir coefficients for the MSSM fields.
The coefficients $c^{h_{\rm u}}_i$ and $c^{h_{\rm d}}_i$
are equal to $c^L_i$.
\label{Tab:bcude}}
\end{table}

\item The experimental error on the
strong coupling constant, that induce large RGE effects,
is sufficiently small for our purposes.
The predictions have been
plotted for $\alpha_3(M_Z) = 0.118$.
To appreciate the impact of this uncertainty,
we have also plotted the predictions of supergravity and
of gauge-mediation with
a very high messenger scale, $\MM=10^{15}\GeV$,
for a somewhat higher value of $\alpha_3$, $\alpha_3(M_Z) = 0.125$,
favoured by unification.

\item In both the minimal models considered, the Higgs masses
are expected to give a non zero contribution to
$$X_Y\equiv \Tr Y_R m_R^2 = (\mhu^2-\mhd^2)+\cdots$$
that, via one-loop RGE effects,
induces a $\sim 5\%$ correction
to the sfermion masses, proportional to their hypercharges $Y$.
We postpone a discussion of this more dangerous correction to the next section,
where we will consider non minimal effects that could
affect the predictions\sys{SuGra} and\sys{GM}.
\end{itemize}
Finally, as pointed out in ref.~\cite{DTW}, we also notice that
the mass ratios under consideration
have a moderate sensivity to the
 `minimal gauge-mediation' parameters, $\MM$, $\eta$,
and can thus be employed for determining their values.

\paragraph{4}
We now discuss how non minimal contributions
to the soft terms could affect the picture described above.
As anticipated, 
we expect that in both scenarios the scalar masses receive a
$\sim 5\%$ correction induced, via RGE effects,
by different boundary conditions for the
Higgs masses, $\mhu\neq\mhd$.
In the case of gauge-mediation models,
such a correction could be much larger.
In fact the U(1)$_Y$ gauge interactions can mediate,
at one-loop, a very large contribution to the squared scalar masses $m_R^2$,
proportional to the hypercharge $Y_R$~\cite{GaugeSoft}:
\begin{equation}\label{eq:YR}
\delta m_R^2 = \eta\eta' Y_R\frac{\alpha_1(\MM)}{4\pi}  M_0^2.
\end{equation}
Since this term is not positive defined, and consequently potentially dangerous,
it is necessary to suppress it by some symmetry.
This happens if the messengers belong to a full degenerate
SU(5) multiplet,
($\Tr Y=0$ over a full SU(5) representation~\cite{GaugeSoft,DG})
or if the messengers appear in a vectorlike $\R\oplus\bar{\R}$ representation
of the gauge group
with a symmetry under $\R\rightleftharpoons \bar{\R}$
($Y_\R + Y_{\bar{\R}}=0$~\cite{GaugeSoft,GMmu}).
Even in these cases, however,
small threshold corrections at the unification scale
that break these symmetries
could result in significant contributions to the sfermion masses.

\begin{figure}[t]\setlength{\unitlength}{1cm}
\begin{center}\begin{picture}(8,8)
\ifMac{\put(0,0){\special{picture 5bis}}}
{\put(0,0){\includegraphics{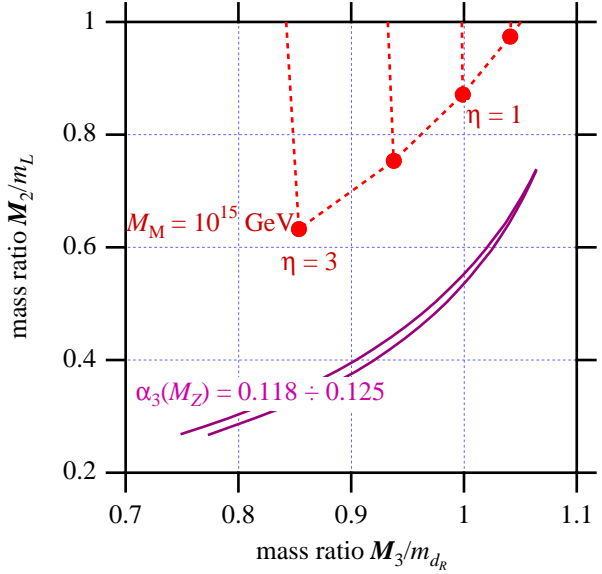}}}
\end{picture}
\caption[1]{\em As in fig.~1,
but in the limit where, due to the correction\eq{YR},
gauge-mediation and unified-supergravity predict an identical
spectrum for the sfermions unified in the $10$ of $\SU(5)$.}
\end{center}\end{figure}

To understand the impact of this correction, let us consider the
{\em two\/} different possible tests
able to discriminate the two scenarios.
For the sake of the argument, let us also assume that
the new correction\eq{YR} mainly affects the lighter slepton masses,
so that the predictions plotted in fig.s~1 and~2 are
shifted vertically by some unknown amount.
However, since the $\tilde{L}$ and $\tilde{e}_R$ sleptons have
hypercharges of opposite sign,
this shift will reduce the separation between
the two scenarios in one of the two tests,
but will increase the separation in the other one.
To be more specific, let us assume that a positive contribution
to the right-handed slepton masses in the gauge mediation scenario
(as favoured by naturalness considerations~\cite{GMFT})
makes to vanish the separation between the continuous and the
dashed lines in fig.~2
(this happens at $\eta'\approx 0.015$).
In this case the separation in the other plot
increases up to the level shown in fig.~3.

In summary, even in presence of a large correction\eq{YR},
one combination of masses allows to
discriminate between the two scenarios
whereas another one allows, in principle, to measure the extra
unknown contribution\eq{YR} proportional to the hypercharges.

\bigskip

To conclude, we now discuss how firm are the
unification relations\sys{SuGra} upon which our analysis is based.

\smallskip

The unification relation between the gaugino masses (\ref{eq:MiGUT})
is easily corrected by the same corrections
that smear the unification prediction
$\alpha_i(\MGUT) = \alpha_5(\MGUT)$.
However, apart for small effects~\cite{SoglieMi},
the corrected unification relation for gaugino masses,
$$M_i(\hbox{just below }\MGUT)\propto \alpha_i(\hbox{just below }\MGUT),$$
(that we have in fact used in our plots in place of\eq{MiGUT})
is very stable (up to few $\%$~\cite{SoglieMi}).
Larger corrections are present only in special circumstances
(if there are many particles
with mass around the unification scale, or if they have much larger soft terms
than the MSSM fields)~\cite{SoglieMi}.

\medskip

The scalar mass relations,\eq{5} and\eq{10}, could be upset
if a SU(5)-breaking field mixes the matter $\bar{5}\oplus10$ fields with
other representations, in such a way that the
light fields do not originate from the same unified
multiplet.
A large mixing is however theoretically disfavoured,
since the additional flavour structures necessary to realize it,
easily produce unacceptable flavour and/or CP violating effects at low energy
(appropriate symmetries could avoid this conclusion).
Barring these mixing effects,
the soft masses of the sfermions of 1$^{\rm st}$ and 2$^{\rm nd}$ generation
with negligible Yukawa couplings
are subject to threshold corrections, which are
again controllable via the corresponding effects
in the gauge couplings.
In particular the corrections
to unification of sfermion masses~(\ref{sys:SuGra}b,c)
are small in models that preserve the successful GUT prediction
of gauge coupling unification.

As a significant example of these facts,
we can consider an interesting model with
large `threshold' corrections with respect to `minimal'
unification, that still gives rise to a successful unification prediction
for the gauge couplings.
In unified models derived from perturbative string theory~\cite{StringGUT},
the adjoint of SU(5) that breaks SU(5) to the standard-model
gauge group has a flat potential,
so that its fragments not eaten by the vector bosons
do not get a mass at the unification scale.
If, for some reason, these chiral supermultiplets with
the same gauge charges of the SM gauge bosons
get mass at an intermediate scale around $10^{13}\GeV$,
the gauge couplings unify
close to the {\em Planck\/} scale~\cite{StringUnific}.
We see that the predictions of this model for the scalar masses
(dotted lines in fig.s~1 and 2)
do not significantly differ from the minimal unification scenario
(continuous lines).

\paragraph{5}
In conclusion we have shown that
the predictions for the sparticle spectrum
of gauge mediated and unified-supergravity scenarios,
as obtained under rather general conditions,
are sufficiently different to
allow the discrimination between
the two scenarios, provided that the sparticle masses,
or rather the appropriate ratios of sparticle masses,
are measured with sufficient precision.
If supersymmetric particles are discovered, it is conceivable that
such measurements will be done at the next generation of accelerators,
supplementing the data obtainable at the hadron collider LHC~\cite{Peskin,LHC},
with those obtainable at an $e^-e^+$ linear collider~\cite{Peskin}.


\newpage

\footnotesize

\begin{thebibliography}{nn}

\bibitem{FVGUT}\art{R. Barbieri and L. Hall}{\PL}{B338}{212}{1994};
\art{S. Dimopoulos and L. Hall}{\PL}{B344}{185}{1995};
\art{R. Barbieri, L. Hall and A. Strumia}{\NP}{B445}{1995}{219}
and {\em \NP \bf B449 \rm(1995) 437}.

\bibitem{FVMSSM} See, e.g.,
\art{S. Bertolini, F. Borzumati, A. Masiero \y G. Ridolfi}{\NP}{B353}{591}{1991};
\art[hep-ph/9610323]{J. Hewett, J.D. Wells}{\PR}{D55}{5549}{1997};
\hepart[hep-ph/9611443]{N.G. Deshpande, B.Dutta \y S. Oh}.

\bibitem{SuGraSoft}
\art{R. Barbieri, S. Ferrara and C.A. Savoy}{\PL}{119B}{343}{1982};
\art{P. Nath, R. Arnowitt and A. Chamseddine}{\PRL}{49}{970}{1982}.

\bibitem{SoglieGUT}
\art{L. Hall, J. Lykken \y S. Weinberg}{\PR}{D27}{2359}{1983};
\art{Y. Kawamura, H. Murayama \y M. Yamaguchi}{\PR}{D51}{1337}{1995};
\art{A. Pomarol \y S. Dimopoulos}{\NP}{B453}{83}{1995}.

\bibitem{RGE} 
\art{L. Ib\'a\~nez, C. Lopez \y C. Mu\~noz}{\NP}{B256}{218}{1985};
\art{A. Boquet, J. Kaplan \y C.A. Savoy}{\NP}{B262}{299}{1985}.

\bibitem{GaugeSoft}
\art{L. Alvarez-Gaume, M. Claudson and M.B. Wise}{\NP}{B207}{96}{1982};
\art{M. Dine \y W. Fishler}{\NP}{B204}{346}{1982}.

\bibitem{DTW}
\art[hep-ph/9609434]{S. Dimopoulos, S. Thomas and J.D. Wells}{\PL}{B357}{573}{1995};

\bibitem{BMPZ}
\art[hep-ph/9609444]{J.A. Bagger, K. Matchev, D.M. Pierce \y R. Zhang}
{\PR}{D55}{3188}{1997}.

\bibitem{GMmu}
\art{G. Dvali, G.F. Giudice and A. Pomarol}{\NP}{B478}{31}{1996}.

\bibitem{Peskin} See, e.g.,
\art[hep-ph/9604339]{M.E. Peskin}
{Prog. Theor. Phys. Suppl.}{123}{507}{1996}
and references therein.

\bibitem{MassRels}
\art{S. Martin \y P. Ramond}{\PR}{D48}{5365}{1993};
\art{H.-C. Cheng \y L.J. Hall}{\PR}{D51}{5289}{1995}.

\bibitem{FTGUT}
\art{S. Dimopoulos and G.F. Giudice}{\PL}{B357}{573}{1995};
\art[hep-ph/9607383]{G. Dvali \y A. Pomarol}{\PRL}{77}{3728}{1996}.

\bibitem{RGE2loop}
\art[hep-ph/9311340]{S.P. Martin \y M. T. Vaghn}{\PR}{D50}{2282}{1994};
\art{Y. Yamada}{\PR}{D50}{3537}{1994};
\art{I. Jack \y D.R.T. Jones}{\PL}{B333}{373}{1994}.

\bibitem{SoglieMZ}
\art{A.B. Lahanas \y K. Tamvakis}{\PL}{B348}{451}{1995};
\art{A. Dedes, A.B. Lahanas \y K. Tamvakis}{\PR}{D53}{3793}{1996}.

\bibitem{1loopSoft}
See, for example,
\hepart[hep-ph/9606211 ]{D.M. Pierce, J.A. Bagger, K. Matchev, R. Zhang}
and references therein.


\bibitem{DG}
\art[hep-ph/9609344]{S. Dimopoulos and G.F. Giudice}{\PL}{B393}{72}{1997}.

\bibitem{GMFT}
\hepart[hep-ph/9611204]{P. Ciafaloni \y A. Strumia};
\hepart[hep-ph/9611243]{G. Bhattacharyya \y A. Romanino}.

\bibitem{SoglieMi}
\art{J. Hisano, H. Murayama \y T. Goto}{\PR}{D49}{1446}{1994}.

\bibitem{StringGUT} See e.g.,
\art{D.C. Lewellen}{\NP}{B337}{61}{1990};
\art{G. Alzadabal, A. Font, L.E. Ib\'a\~nez \y A.M Uranga}{\NP}{B452}{3}{1995};
\hepart[hep-th/9610106]{Z. Kakushadze \y S.-H.H. Tye}.

\bibitem{StringUnific}
\art{C. Bachas, C. Fabre \y T. Yanagida}{\PL}{B370}{49}{1996}.

\bibitem{LHC}
\art[hep-ph/9610544]{I. Hinchliffe, F.E. Paige, M.D. Shapiro,
J. S\"oderqvist \y W. Yao}{\PR}{D55}{5520}{1997}.

\end{thebibliography}
\end{document}
\\
Title: Distinguishing gauge-mediated from unified-supergravity spectra
Authors: Alessandro Strumia
Comments: 6 pages
Report-no: IFUP-TH 18/97
\\
We show that gauge-mediation and unified-supergravity
give sufficiently firm and different predictions for the spectrum
of supersymmetric particles
to make it possible to discriminate the two scenarios
even if the messenger mass is close to the unification scale.
\\